\documentclass{article}

\usepackage{mathptmx}       
\usepackage{helvet}         
\usepackage{courier}        
\usepackage{type1cm}       

\usepackage{makeidx}         
\usepackage{graphicx}        

\usepackage{multicol}        
\usepackage[bottom]{footmisc}

\makeindex            

\usepackage{amsmath}
\usepackage{amssymb}
\usepackage{caption}
\usepackage{subcaption}
\usepackage[labelfont=bf]{caption}

\begin{document}	
	\title{Inferring differentiation order in adaptive immune responses from population level data}

	\author{Alexander S. Miles\thanks{Hamilton Institute, Maynooth University, Ireland}, Philip D. Hodgkin\thanks{ Walter and Eliza Hall Institute, Royal Melbourne Hospital, Parkville,
                Australia and the Department of Medical Biology, The University of
                Melbourne, Parkville, Australia}, and Ken R. Duffy\thanks{Hamilton Institute, Maynooth University, Ireland. E-mail: Corresponding ken.duffy@nuim.ie}}

	\maketitle
	\abstract{A hallmark of the adaptive immune response is the
	proliferation of pathogen-specific lymphocytes that leave
	in their wake a long lived population of cells that provide
	lasting immunity. A subject of ongoing investigation is
	when during an adaptive immune response those memory cells
	are produced. In two ground-breaking studies, Buchholz et
	al. (Science, 2013) and Gerlach et al. (Science, 2013)
	employed experimental methods that allowed identification
	of offspring from individual lymphocytes \textit{in vivo},
	which we call clonal data, at a single time point. Through the
	development, application and fitting of a mathematical model,
	Buchholz et al. (Science, 2013) concluded that, if memory
	is produced during the expansion phase, memory cell precursors are
	made before the effector cells that clear the original
	pathogen. We sought to determine the general validity and
	power of the modeling approach introduced in Buchholz et
	al. (Science, 2013) for quickly evaluating differentiation
	networks by adapting it to make it suitable for drawing
	inferences from more readily available non-clonal phenotypic proportion
	time-courses. We
	first established the method drew consistent deductions
	when fit to the non-clonal data in Buchholz et al. (Science,
	2013) itself. We fit a variant of the model to data reported
	in Badovinac et al. (Journal of Immunology, 2007), Schlub
	et al. (Immunology and Cell Biology, 2010), and Kinjo et
	al. (Nature Communications, 2015) with necessary simplifications to match different reported data in these papers. The deduction from the
	model was consistent with that in Buchholz et al. (Science, 2013),
	albeit with questionable parameterizations. An alternative
	possibility, supported by the data in Kinjo et al. (Nature
	Communications, 2015), is that memory precursors are created after the
	expansion phase, which is a deduction not possible from the
	mathematical methods provided in Buchholz et al. (Science,
	2013). This investigation further supports the value of the
	approach, though we indicate where published experimental
	findings run contrary to assumptions underlying the model,
	which may impact the strength of inferences, and what an
	alternate model more consistent with recent data might be.}

	\section{Introduction} \label{sec:Intro} 

	During an adaptive immune response, a population of naive
	lymphocytes becomes heterogeneous, containing a population
	of effector cells that clear the pathogen and then die
	during the contraction phase, and a population of memory
	cells that are of a longer lineage. For CD8\textsuperscript{+}
	T cells, many theories have been proposed for the dynamics of
	this differentiation. The three primarily championed ones
	are \cite{ModAndCont,lemaitre2013phenotypic,buchholz2016t}: 
		\begin{itemize}
		\item \textbf{Linear Effector First Model:} Naive
		cells differentiate into proliferating effector
		cells. Later in the immune response, these cells
		differentiate into memory cells or die 
		\cite{Opferman1745,Eff1stSup1,Wherry2003,Kinjyo2015}.
		\item \textbf{Linear Memory First Model:} Naive
		cells differentiate into memory precursor cells that then
		proliferate, with some differentiating into
		effector cells. Some time after the expansion phase is completed
		the effector cells die \cite{Buchholz630}.
		\item\textbf{Bifurcation Model:} Each naive cell
		has the potential to divide into either effector
		or memory precursor subsets which then continue to proliferate
		inheriting their phenotype, and effector cells die
		after the expansion phase while memory cells remain
		\cite{Chang1687,king2012t}.
	\end{itemize} 
	These three are non-exhaustive, high-level summaries of a
	more diverse and subtle set of theories. Provision of
	a simple classification is further complicated by phenotypic
	characterization of T cells where there is an overlap between
	memory and effector functionality and phenotypic markers,
	sometimes resulting in a sub-category that is referred to
	as effector memory cells (TEMp, the p standing for precursor)
	to distinguish them from effector (TEF) and central memory
	cells (TCMp).

	The Linear Effector First Model has a long history and could
	be considered the traditional view \cite{Opferman1745,ModAndCont}.
	In 2007, the authors of \cite{Chang1687} proposed asymmetric
	cell division to explain heterogeneity with CD8\textsuperscript{+}
	T cells, supporting the Bifurcation Model, which won't be
	considered further here. In 2013, two landmark papers were
	published \cite{Buchholz630,Gerlach635} that further
	addressed this question.  Both used novel methodologies to
	determine clonal relationships of cells. In addition, due
	to the more quantitative nature of their assay, the authors
	of \cite{Buchholz630} developed a methodology to fit a
	memoryless multi-type Bellman-Harris process model \cite{BHprocess}
	to clonal summary statistics of CD8\textsuperscript{+}
	T cell phenotypes, finding that a quantitative realization
	of the Linear Memory First Model was the best explanation
	of their data.

	Let us introduce some notation in order to distinguish
	between the available data sources. For an experiment
	initiated at time $t=0$ with $n$ pathogen specific cells,
	let $c^f_i(t)$ denote the number offspring that are of
	phenotype $f\in\{1,\ldots,F\}$ from an initial transferred
	naive cell $i$ at time $t$ post infection. The method
	used in \cite{Buchholz630} was effectively unique in
	determining the number of cells of each phenotype that were
	descendent from an initially naive cell, $c^f_i(t)$, for
	each $i$ and all $f\in\{\text{TEF, TEMp, TCMp}\}$  at $t=$
	day eight post infection. We shall refer to this data as
	being at a clonal level.

	A more common experimental set-up
	\cite{Buchholz630,Badovinac2007,SchlubMain,Kinjyo2015}
	provides the fraction of all offspring that are of each 
	phenotype, $f$, at a given time $t$,
	\begin{align*}
	\rho^f(t) = \frac{\sum_{i=1}^n c^f_i(t)}{\sum^F_{j=1}\sum_{i=1}^n c^j_i(t)},
	\end{align*}
	with clonal contributions being indistinguishable. Experiments
	recording this information are typically performed as a
	time-course, with cells harvested at distinct days post
	infection. We shall refer to it as cohort or population
	level data.

	In the present paper, we describe deductions made by taking
	the modeling framework described in \cite{Buchholz630} and
	adapting it to make it suitable for fitting to cohort data.
	Our aim was to explore the general validity and power of
	the method for quickly evaluating differentiation networks
	from simple cohort time series as, unlike clonal data, those
	data are more readily available. We used the method to
	interrogate data reported by other authors
	\cite{Badovinac2007,SchlubMain,Kinjyo2015},
	that employ similar experimental systems to that described
	in \cite{Buchholz630}.

		A high-level summary of the content of this paper is:
		\begin{enumerate}
			\item
			To adapt the method in \cite{Buchholz630} for
			fitting to  
			non-clonal data, we introduce a distinct
			objective function. On comparing the best
			parameter fits to six representative models
			of the 304 considered in the original paper
			to the time-course cohort data reported in
			\cite{Buchholz630} itself, the finding was
			the same as for the clonal data: that if
			memory is made during the expansion phase,
			then the Linear Memory First Model is the
			best. Moreover, the best-fit parameterization
			from the cohort data was quantitatively
			similar to that when fit to clonal data,
			suggesting consistency of the method.
			\item
			Due to data limitations where fewer phenotypes were
			identified, for the papers \cite{Badovinac2007,SchlubMain,Kinjyo2015} we
			fit only to two representative differentiation
			structures: Linear Effector First and Linear
			Memory First. Naive application of the
			methodology to the entire time-course cohort data taken from
			\cite{Kinjyo2015} supported the Linear Effector First Model
			over the Linear Memory First Model.
			A more sophisticated application was performed,
			noting that the number of
			antigen-specific T cells adoptively transferred
			in the experiments reported in \cite{Kinjyo2015}
			was significantly higher than in
			\cite{Buchholz630}, resulting in a curtailed
			expansion phase. Limiting model fitting to expansion
			phase only, data from blood and spleen in
			four papers, \cite{Badovinac2007,SchlubMain,
			Buchholz630, Kinjyo2015} all support a
			Linear Memory First Model in favor of a
			Linear Effector First Model. Best fit model
			parameterizations, however, suggest the
			model is struggling to explain this data.
			\item
			Additional data reported in
			\cite{Kinjyo2015} suggests that the \textit{a
			priori} assumption underlying the mathematical
			analysis that memory is created during the
			expansion phase could be questionable.
			\item
			While the evidence presented here further
			supports the utility of the method introduced
			in \cite{Buchholz630}, we indicate some of
			the discrepancies between its model assumptions
			and published data, suggesting an alternate
			model structures that are more consistent
			with them, which might be of use in future.
		\end{enumerate}

	\section{Experiments and results from relevant papers}
\begin{figure}[h]
		\begin{subfigure}[b]{1\textwidth}
			\caption{.}
			\includegraphics[width=\textwidth]{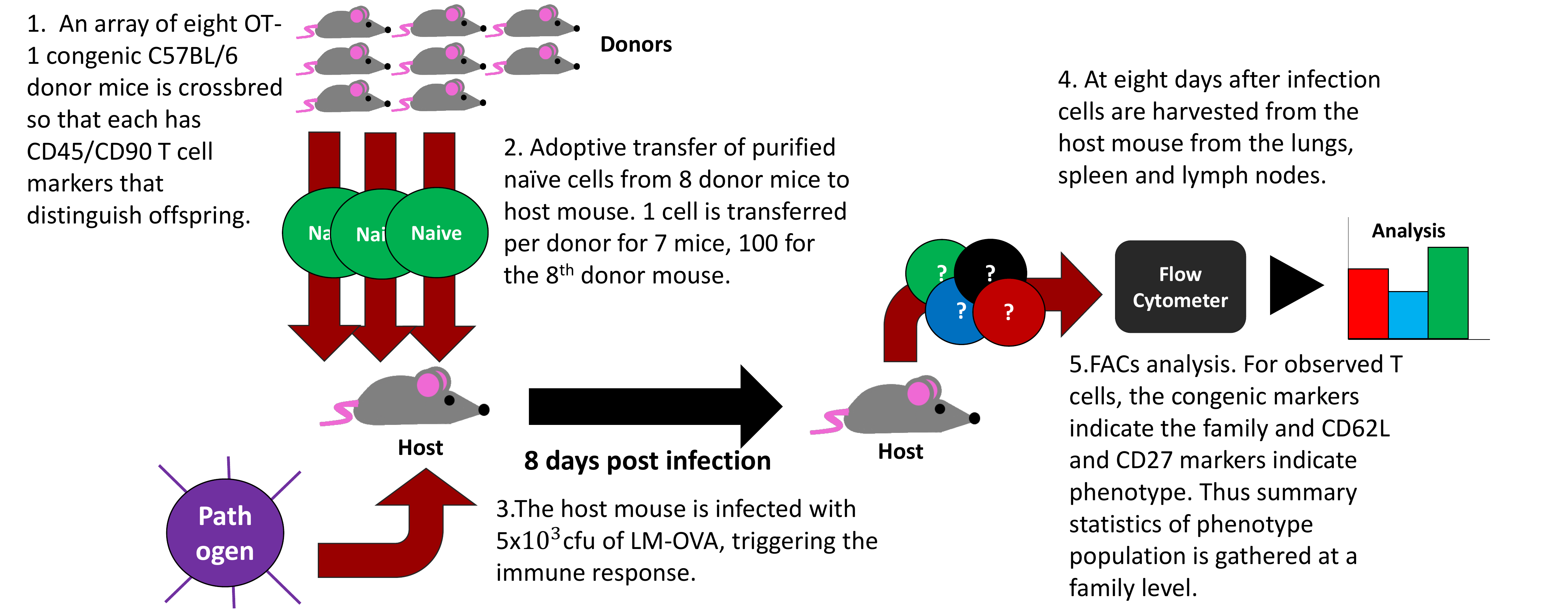}\label{fig:1a}
		\end{subfigure}
		\begin{subfigure}[b]{1\textwidth}
			\caption{.}
			\includegraphics[width=\textwidth]{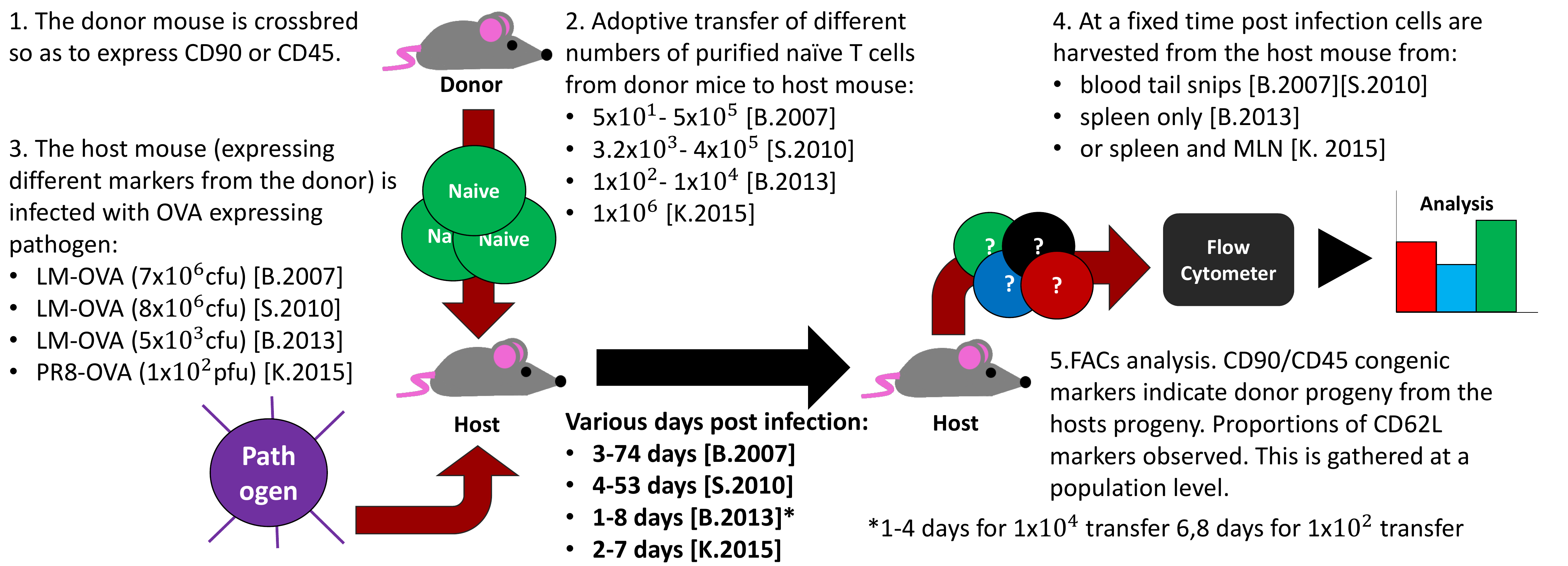}\label{fig:1b}
		\end{subfigure}
		\caption{\textbf{Summary adoptive
		transfer experiments described in
		\cite{Badovinac2007,SchlubMain,Buchholz630,Kinjyo2015}.} 
		(A)
		Experiments described in \cite{Buchholz630} determine
		the number of cells per OT-1 CD8+ T cell clone 
		at day eight. (B) All four papers describe similar
		OT-1 CD8+ T cell adoptive transfer experiments,
		where cohort data is reported for a time-course of
		days post infection. The number of cells adoptively
		transferred varies between papers.}
	\end{figure}

	\textbf{Clonal level experiments \cite{Buchholz630}}:
	The authors performed innovative OT-1 CD8+ T cell adoptive transfer
	experiments where cell lineage could be determined. Donor
	mice were engineered to express both CD45 and CD90 markers
	homozygously, heterozygously or not at all, allowing for
	nine combinations of markers in total, Fig. \ref{fig:1a}.
	C57BL/6 host mice were chosen from the double heterogeneous
	marker pool.  Recipient mice received a mix of naive OT-1
	CD8 T cells that contained one cell from each of the seven
	uniquely congenically marked donor mice together with 100 cells
	from one further uniquely marked mouse. As lineage was
	determined by cell surface markers, clonal cell numbers
	could be quantitatively evaluated by FACS.  The cellular
	barcoding approach used in the twinned paper \cite{Gerlach635}
	was based on DNA tags and their processing via PCR makes
	that data less suitable for quantitative modeling analysis.

\begin{figure}[h]
		\begin{subfigure}[b]{1\textwidth}
		\caption{.}
		\includegraphics[width=\textwidth]{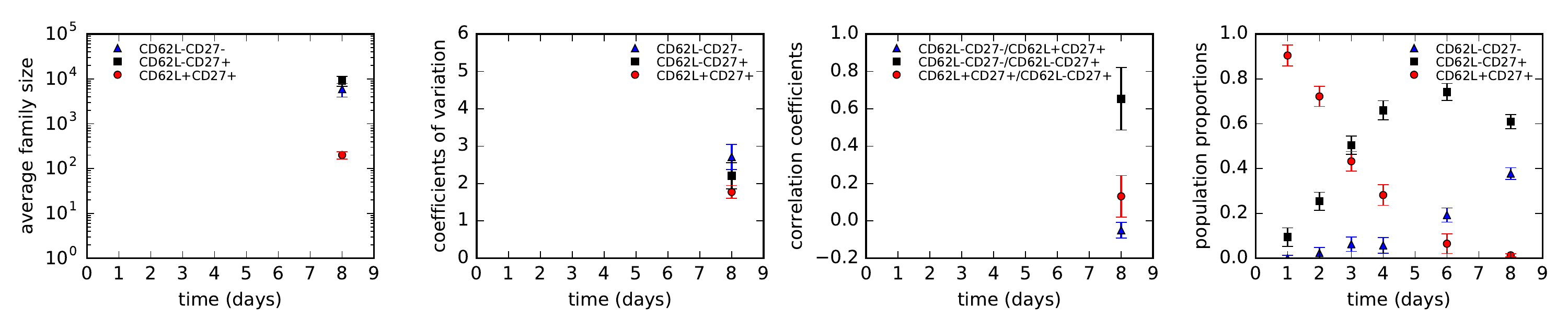}\label{fig:2a}
		\end{subfigure}
		\begin{subfigure}[b]{1\textwidth}
		\caption{.}
		\includegraphics[width=\textwidth]{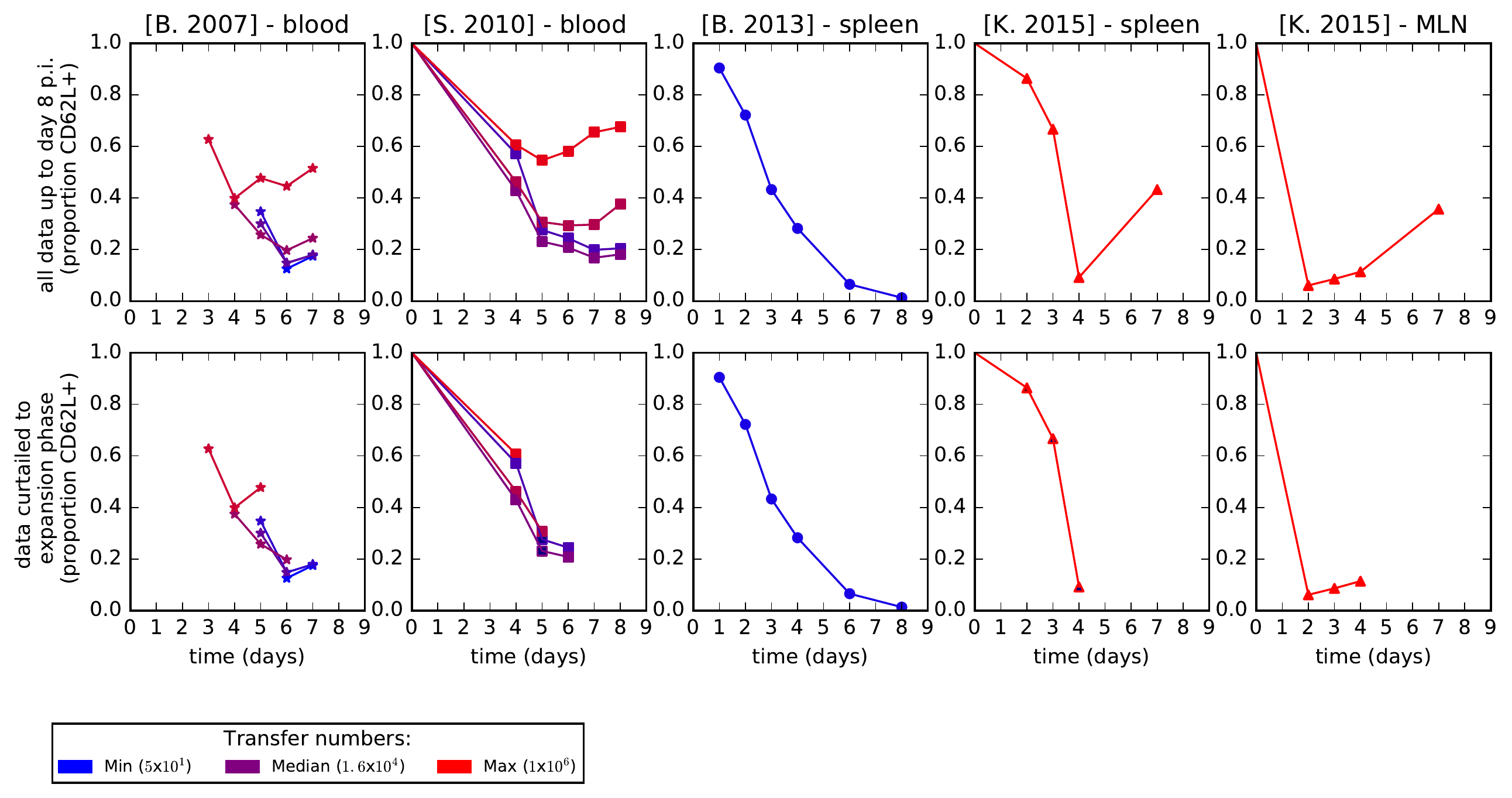}\label{fig:2b}
		\end{subfigure}
		\caption{\textbf{Experimental data from 
		\cite{Badovinac2007, SchlubMain,Buchholz630,Kinjyo2015}.} Data
		manually extracted from graphs in those papers. (A, left
		three graphs) Results from the clonal experiment
		described in \cite{Buchholz630}, showing average
		family size, coefficients of variation and correlations
		for different phenotypes populations. (A, rightmost
		graph) Proportional phenotype results from the
		cohort experiment. (B) Comparing the
		CD62L\textsuperscript{+} proportions from the four
		papers reporting cohort data. The top row shows all
		reported data up to day eight post infection. The
		bottom row shows the data when curtailed to the
		estimated expansion phase, which is shortened due to
		high adoptive transfer cell numbers. }
		\label{fig:Results} 
	\end{figure}

	In the experiments reported in \cite{Buchholz630}, the host
	was infected with $5 \times 10^3$ Listeria monocytogenes,
	a pathogenic bacteria, expressing OVA (LM-OVA).  At day
	eight post infection, cells were
	taken from the spleen, lymph nodes and lungs.
	The cells were sorted into three phenotypes: 
	\begin{itemize} 
	\item \textbf{TCMp}: CD62L\textsuperscript{+}CD27\textsuperscript{+} phenotype.
	\item \textbf{TEMp}: 	CD62L\textsuperscript{-}CD27\textsuperscript{+} phenotype.
	\item \textbf{TEF}: 	 CD62L\textsuperscript{-}CD27\textsuperscript{-} phenotype.
	\end{itemize} 
	Summary statistics from the data collected for these phenotypically defined populations
	were reported: the average cell count per family, coefficients
	of variation and correlation coefficients between families
	for the three phenotypes, Fig. \ref{fig:2a}.

	\textbf{Cohort level experiments, \cite{Badovinac2007,SchlubMain,Buchholz630,Kinjyo2015}}
	Each of these four papers also performed OT-1 CD8+ T cell adoptive
	transfer experiments, Fig. \ref{fig:1b}. In these, cells
	were not tracked at a clonal level, $\{c^f_i(t)\}$, but at a cohort level
	$\{\rho^f(t)\}$,
	via expression of either CD45 or CD90. The host mouse was
	infected with a pathogen expressing OVA; PR8-OVA in
	\cite{Kinjyo2015} and LM-OVA in the other papers. The
	reported experiments have cells taken from different
	organs, spleen, or mesenteric lymph nodes (MLN), at
	a range of times post infections. In addition,
	\cite{Badovinac2007,SchlubMain} performed experiments with
	varying number of transferred OT-1 T cells. All of those
	papers report the percentage of cells that were
	CD62L\textsuperscript{+} at harvest times.

For cohort spleen and blood data reported in all four papers, at
early stages post infection all experiments show an initial large
CD62L\textsuperscript{+} population, Fig. \ref{fig:2b}, which
the authors of \cite{Buchholz630} associate this with an early
memory precursor population while the authors of \cite{Kinjyo2015} associate
it with cells still being naive. All papers show a decrease in
this initial CD62L\textsuperscript{+} proportion, which both
\cite{Kinjyo2015} and \cite{Buchholz630} explain is due to the effector
pool increasing in size. The paper \cite{Kinjyo2015} reports an
increase in the percentage of CD62L\textsuperscript{+} by day seven
and explains this as cells differentiating into memory cells, whereas
the data in \cite{Buchholz630} continues to show a decrease by day
seven.

For experiments with a high number of adoptively transferred cells,
\cite{SchlubMain} and \cite{Badovinac2007} report an earlier expansion
peak and a shorter expansion phase. Experiments transferring
less than 500 cells had a peak at day seven, while for experiments
transferring $5 \times 10^5$ cells the peak immune response occurred
at day five. Curtailing the experimental data with higher numbers
of adoptively transferred cells to its expansion phase (a method for estimating this time we describe later in the paper) removes much
of the later upturn in CD62L proportions, Fig. \ref{fig:2b}.
Having transferred $5 \times 10^5$ cells, the paper \cite{Badovinac2007} estimates the family size increase to be 40 -- 400 fold on average at the peak, compared to
the estimated 400,000 fold increase on average for the experiment
transferring $50$ cells.

	\section{The mathematical model, its adaptation \& fit
	to cohort data reported in \cite{Buchholz630}}

	The authors of \cite{Buchholz630} introduced a collection
	of memoryless multi-type Bellman-Harris stochastic processes
	\cite{BHprocess} to model the expansion phase of an adaptive
	immune response based on the following assumptions.
	\begin{itemize}
		\item There are four types of cell: naive, TCMp, TEMp, and TEF. 
		\item There is no cell death during the expansion phase described by the model.
		\item For each cell type, the lifetime of each cell
		is an independent and identically exponentially
		distributed random variable with a cell-type dependent
		mean.
		\item At the end of its lifetime, a naive cell differentiates
		 into one of the other cell types.
		\item TCMp, TEMp, and TEF cells can divide or
		differentiate, with the each cell's fate determined
		probabilistically and independently based on a cell-type
		dependent parameterization.
		\item Possible differentiation between cell-types is known as a
		path. Different model types have a different
		combination of paths between cell types, which can
		be uni- or bi-directional.
		\item There must be a direct or indirect path from
		the naive cell to each cell-type.
		\item When the combinatorics are all done, the
		collection of possible paths describes 304 different
		models with between seven and 12 parameters.
	\end{itemize}
	For a given differentiation network and parameterization,
	the authors of \cite{Buchholz630} employ a $\chi^2$ distance
	as a measure of a quality of fit between expected summary
	statistics and observed summary statistics. Denote the nine
	observed clonal summary statistics per-phenotype of mean family
	size, coefficient of variation and Pearson correlation
	between the phenotypes as $\hat{x} =(\hat{x}_1,...,\hat{x}_9)
	\in \mathbb{R}^9$, and their equivalent expected values for
	a given model plus parameterization, $\theta$, as
	$\vec{x}(\theta)\in \mathbb{R}^9$. With experimental
	uncertainty of each observation, $\hat{x}_i$, estimated by
	bootstrap approximation as $\hat{\sigma}_i$, the $\chi^2$ distance
	between models and observations is computed to be
        \begin{align*}
        d_1(\hat{x},\vec{x}(\theta)) = \sum_{i=1}^9\frac{(\hat{x_i} - x_i(\theta))^2}{\hat{\sigma}_i^2}.
        \end{align*}
	For each of the 304 possible models, the best-fit
	parameterization, $\theta^*=\arg\inf_{\theta}
	d_1(\hat{x},\vec{x}(\theta))$, was numerically identified
	and more involved models were penalised via Akaike Information
	Criterion (AIC). One of the two best fits
	over all 304 possible differentiation networks was a linear differentiation pathway with memory cells
	appearing before effector cells (the second model was as the first, but with an additional 10\% chance for naive to skip the TCMp stage of the path).

	We wished to determine the method's consistency when fitting
	the model class described in \cite{Buchholz630} to the
	time-course cohort data $\{\rho^f(t): t=\text{day }1,2,3,4,6,8
	\}$, described in that paper rather than to the clonal data
	$\{c^f_i(t):t=\text{day } 8\}$ as the authors had originally
	done. To do so, we replaced the clonal summary statistics
	with $\hat{y} = (\hat{y}_1,...,\hat{y}_{19})$ where $\hat{y}_1$
	is the average family size at the peak and $y_2,...,y_{19}$
	are the 18 measured proportions of phenotypes observed at
	different times post infection. We denote $\hat{\sigma_i}$
	as the sample error of the statistic $\hat{y}_i$. In the
	case of the average family size, the SEM was approximated
	by summing together the SEMs of the phenotypic components.
	Thus, we use the formulae
        \begin{align*}
        d_2 (\hat{y},\vec{y}(\theta)) = \sum_{i=1}^{19}\frac{(\hat{y_i} - y_i(\theta))^2}{\hat{\sigma}_i^2}.
        \end{align*}
	to measure the distance between expected and observed cohort 
	statistics.

\begin{figure}[h]
		\begin{subfigure}[t]{\textwidth} \caption{.}
		\includegraphics[width=\textwidth]{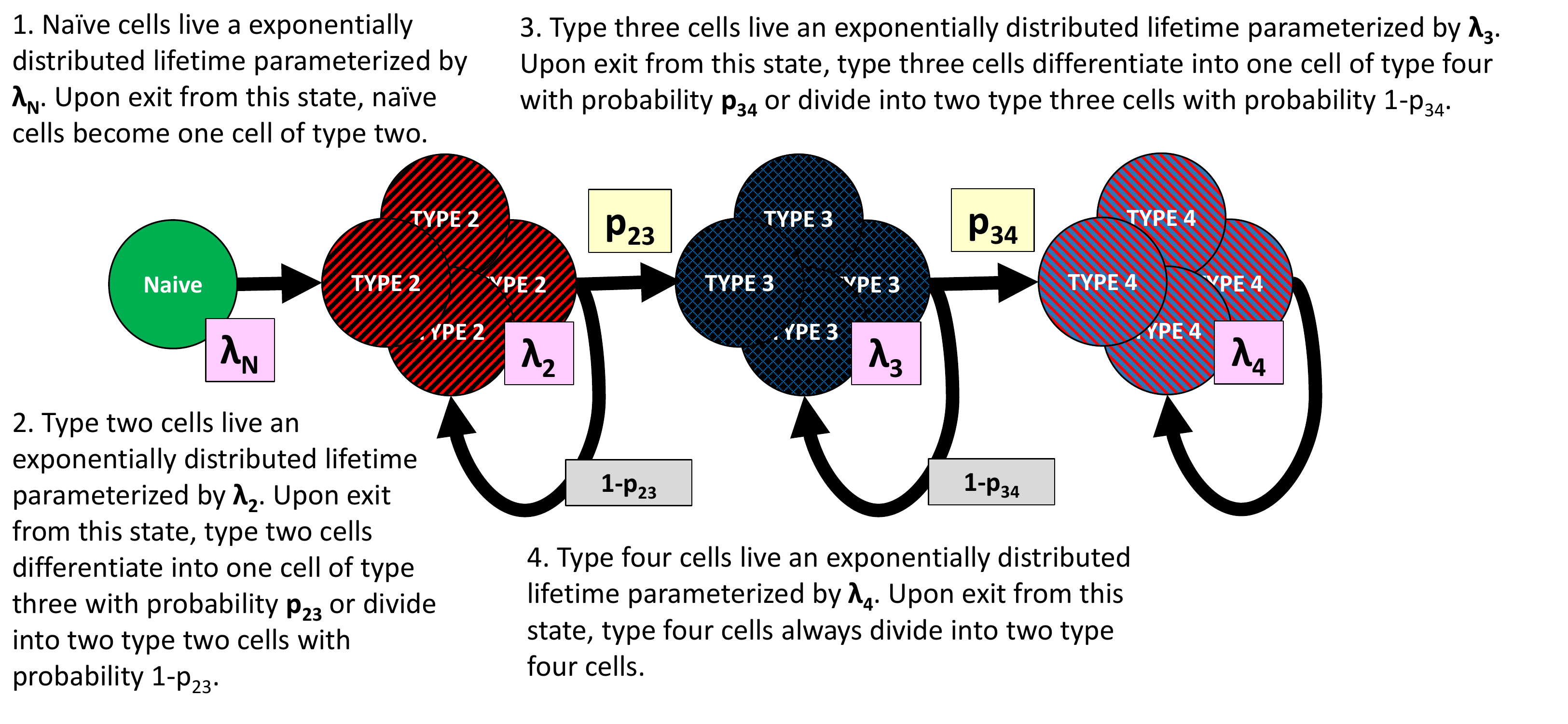}\label{fig:3a}
		\end{subfigure} \begin{subfigure}[t]{\textwidth}
		\caption{.}
		\includegraphics[width=\textwidth]{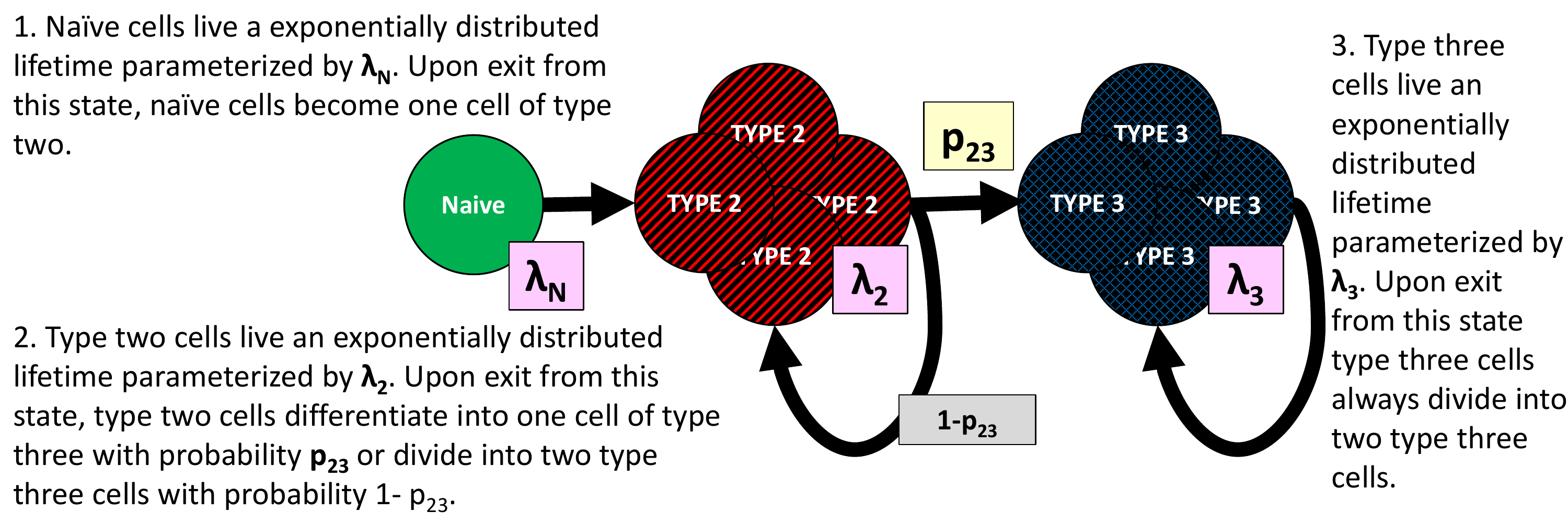}\label{fig:3b}
		\end{subfigure} \caption{\textbf{Linear
		differentiation models considered here.} (A) These
		linear models are a subset of the 304 multi-type Bellman-Harris
		processes modeled by \cite{Buchholz630}. Each of
		the six models has six parameters, $(\lambda_N,\lambda_2,\lambda_3,\lambda_4,p_{23},p_{34})$.
		(B) Simplified linear models, with the observed
		phenotypes reduced from three
		(CD62L\textsuperscript{+}CD27\textsuperscript{+},
		CD62L\textsuperscript{+}CD27\textsuperscript{-},
		CD62L\textsuperscript{-}CD27\textsuperscript{-})
		to two (CD62L\textsuperscript{+}, CD62L\textsuperscript{-})
		and cell types reduced from four to three. The
		simplified model has four parameters,
		$(\lambda_N,\lambda_2,\lambda_3,p_{23})$.}
		\label{Fig:LinearModels}
	\end{figure}

	We numerically identified the best fit parameterization of the model for six
	representative differentiation pathways, Fig. \ref{fig:3a},
	of the 304 considered in \cite{Buchholz630}. These six were chosen since they include one of the two best fit models and were sufficient to recreate the most significant results reported in \cite{Buchholz630}. We label the
	six models by:

	\begin{itemize}
		\item \textbf{Model 1}: Naive $\rightarrow$
		CD62L\textsuperscript{+}CD27\textsuperscript{+}
		$\rightarrow$
		CD62L\textsuperscript{-}CD27\textsuperscript{+}
		$\rightarrow$
		CD62L\textsuperscript{-}CD27\textsuperscript{-}.
		\item \textbf{Model 2}: Naive $\rightarrow$
		CD62L\textsuperscript{+}CD27\textsuperscript{+}
		$\rightarrow$
		CD62L\textsuperscript{-}CD27\textsuperscript{-}
		$\rightarrow$
		CD62L\textsuperscript{-}CD27\textsuperscript{+}.
		\item \textbf{Model 3}: Naive $\rightarrow$
		CD62L\textsuperscript{+}CD27\textsuperscript{-}
		$\rightarrow$
		CD62L\textsuperscript{+}CD27\textsuperscript{+}
		$\rightarrow$
		CD62L\textsuperscript{-}CD27\textsuperscript{-}.
		\item \textbf{Model 4}: Naive $\rightarrow$
		CD62L\textsuperscript{+}CD27\textsuperscript{-}
		$\rightarrow$
		CD62L\textsuperscript{-}CD27\textsuperscript{-}
		$\rightarrow$
		CD62L\textsuperscript{+}CD27\textsuperscript{+}.
		\item \textbf{Model 5}: Naive $\rightarrow$
		CD62L\textsuperscript{-}CD27\textsuperscript{-}
		$\rightarrow$
		CD62L\textsuperscript{+}CD27\textsuperscript{+}
		$\rightarrow$
		CD62L\textsuperscript{+}CD27\textsuperscript{-}.
		\item \textbf{Model 6}: Naive $\rightarrow$
		CD62L\textsuperscript{-}CD27\textsuperscript{-}
		$\rightarrow$
		CD62L\textsuperscript{-}CD27\textsuperscript{+}
		$\rightarrow$
		CD62L\textsuperscript{+}CD27\textsuperscript{+}.
	\end{itemize} 
	Model 1 is one of the two original best fitting models
	reported in \cite{Buchholz630}.  As these models all have
	the same number of parameters, we do not need to consider
	penalization by AIC.  Due to less extensive data in
	\cite{Badovinac2007,SchlubMain,Kinjyo2015}, later we will
	reduce the set model further to consider just two competing
	models, Memory (CD62L\textsuperscript{+}) First and Effector
	(CD62L\textsuperscript{-}) First, Fig. \ref{fig:3b}.

	The model in \cite{Buchholz630} has
	four cell types, but only reports data for three phenotypes and 
	no statistics for naive cells are presented. This could be
	interpreted two ways, either naive cells have been excluded
	from the data, or they were included in the count of one of
	the other phenotypes. We note that \cite{Buchholz630} define
	naive cells as CD62L\textsuperscript{+} CD27\textsuperscript{+}
	(\cite{Buchholz630} Supplementary Fig. 3), but with a
	different gating strategy than used for the classification
	of other cell types (\cite{Buchholz630} Supplementary Fig.
	22). For the analysis here, we assume that naive cells
	are included in the TCMp count reported in \cite{Buchholz630}.

\begin{figure}[b]
		\begin{subfigure}[t]{0.5\textwidth}
			\caption{.}
			\includegraphics[width=\textwidth]{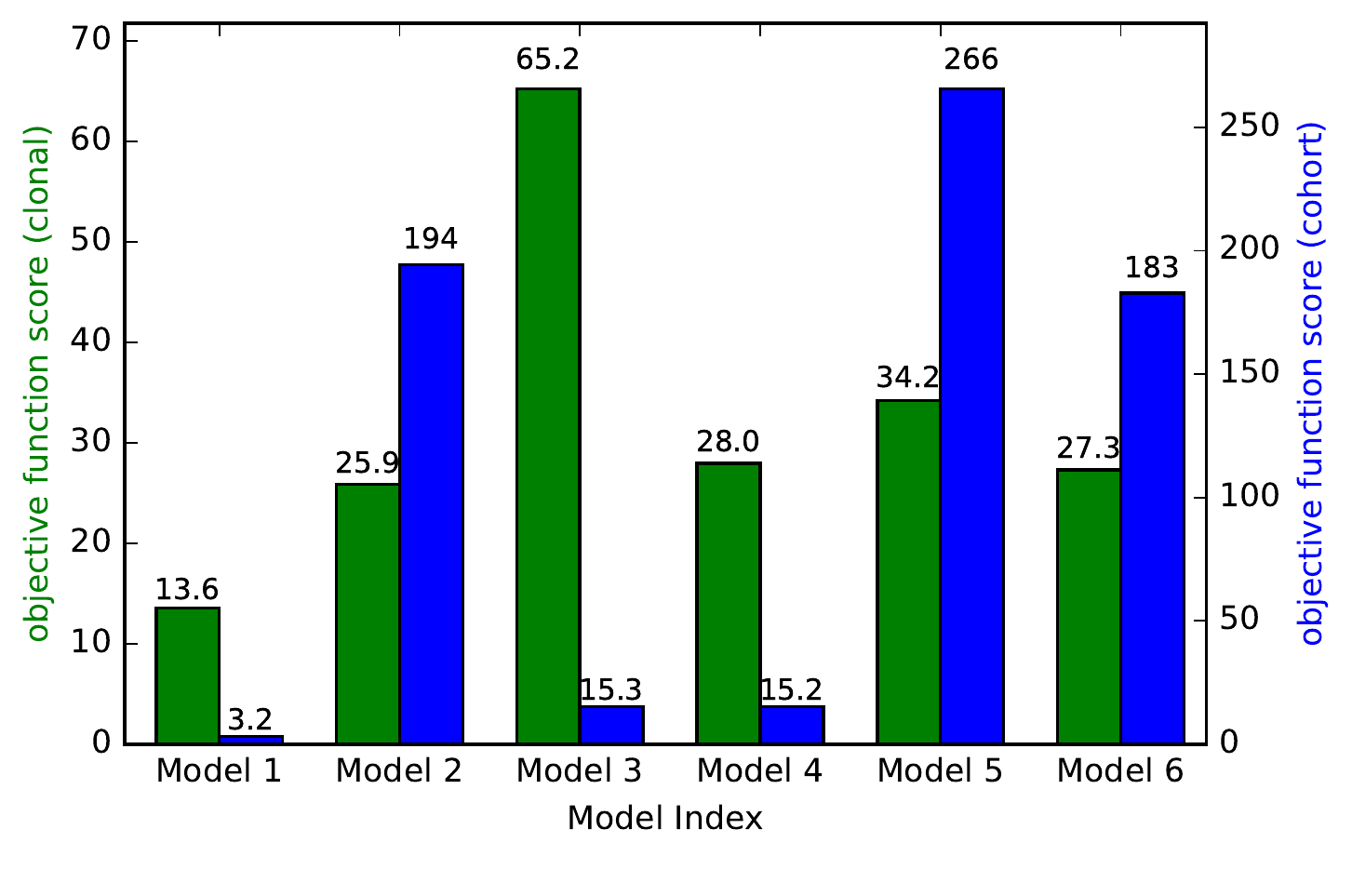}
			\label{fig:4a}
		\end{subfigure} \begin{subfigure}[t]{0.5\textwidth}
			\caption{.}
			\includegraphics[width=0.92\textwidth]{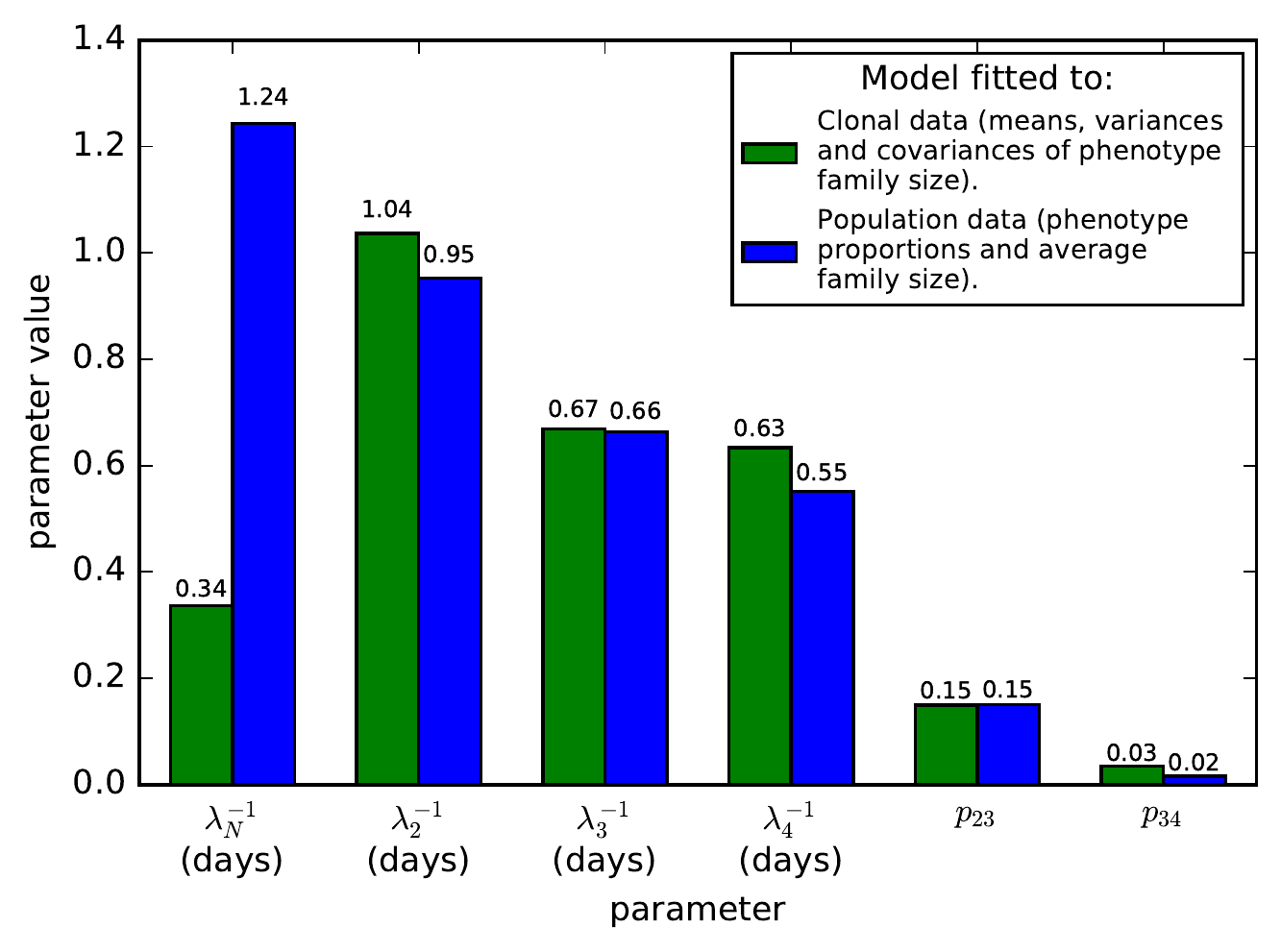}\label{fig:4b}
		\end{subfigure} \caption{\textbf{Results from fitting
		the linear models to clonal data, as in the original method,
		or cohort data from \cite{Buchholz630}.} (A) Objective function
		values for the best fitting parameterization of each
		model, where smaller values indicate a better
		fit. (B) Best fitting parameterization for Model 1, where
		$\lambda_2$ corresponds to CD62L\textsuperscript{+}CD27\textsuperscript{+}
		(TCMp),
		$\lambda_3$ to CD62L\textsuperscript{-}CD27\textsuperscript{+}
		(TEMp), 
		$\lambda_4$ to CD62L\textsuperscript{-}CD27\textsuperscript{-}
		(TEF), $p_{23}$ to the probability that
		a TCMp cell becomes a TEMp cell and $p_{34}$ to the probability that
		a TEMp cell becomes a TEF cell.}
	\end{figure}

	When applied to the time-course cohort data reported in
	\cite{Buchholz630}, we found that Model 1, Memory First,
	to be the best fit, consistent with the original deduction
	from fitting to clonal data, Fig. \ref{fig:4a}. Notably,
	the parameterization for the best fitting model was remarkably
	similar for five parameters out of six, as shown in Fig.
	\ref{fig:4b}. Fitting to either data-set is consistent with
	an increase in the division rate as cells differentiate
	from memory to effector cells, as also reported in
	\cite{Buchholz630}. The difference in the parameter $\lambda_N$,
	which parameterizes the lifetime of naive cells, could be
	explained by the fact that the adjusted method compares the
	sum of proportions of naive and TCMp cells to one statistic,
	the proportion of CD62L\textsuperscript{+}CD27\textsuperscript{+}
	cells, and is not able to discriminate between their separate
	contributions. These results suggest that, for this model,
	fitting to time-course cohort data with an average family
	size is a reasonable alternative to fitting to clonal data
	of mean count of cells, variances and covariances.

	\section{Adaptation and application to cohort data from \cite{Badovinac2007,SchlubMain,Kinjyo2015}} 
	\label{Sec:FurtherAdap} 

	\textbf{Adaptation.}
	As data from other papers we wish to analyze,
	\cite{Badovinac2007,SchlubMain,Kinjyo2015}, did not include all of 
	the information as \cite{Buchholz630} it was necessary to use a modified scheme. In some cases, average family size or
	SEM information was not given. All four papers
	provide the proportion of CD62L\textsuperscript{+} cells,
	but some do not report CD27 data. Thus we define a new set
	of models that do not distinguish between CD27\textsuperscript{-}
	and CD27\textsuperscript{+} cells so that we can fit the
	observed statistics without this information. We change the
	assumption that there are four underlying cell types: naive,
	TCMp, TEMp and TEF used in \cite{Buchholz630}, to three
	cell types:  naive, memory and effector cells.
	Thus the phenotypes are:
	\begin{itemize}
	\item \textbf{Naive and memory cells}: 	CD62L\textsuperscript{+} phenotype.
	\item \textbf{Effector cells}: 	 		CD62L\textsuperscript{-} phenotype.
	\end{itemize}
	 
	We label the two versions of these models as:
	\begin{itemize}
		\item \textbf{Linear Memory First Model} differentiation
		path: Naive $\rightarrow$ CD62L\textsuperscript{+}
		$\rightarrow$ CD62L\textsuperscript{-}.
		 \item
		\textbf{Linear Effector First Model} differentiation
		path: Naive $\rightarrow$ CD62L\textsuperscript{-}
		$\rightarrow$ CD62L\textsuperscript{+}.
	\end{itemize} 

\begin{figure}[h]
		\begin{subfigure}[t]{0.5\textwidth} \caption{.}
		\includegraphics[width=0.97\textwidth]{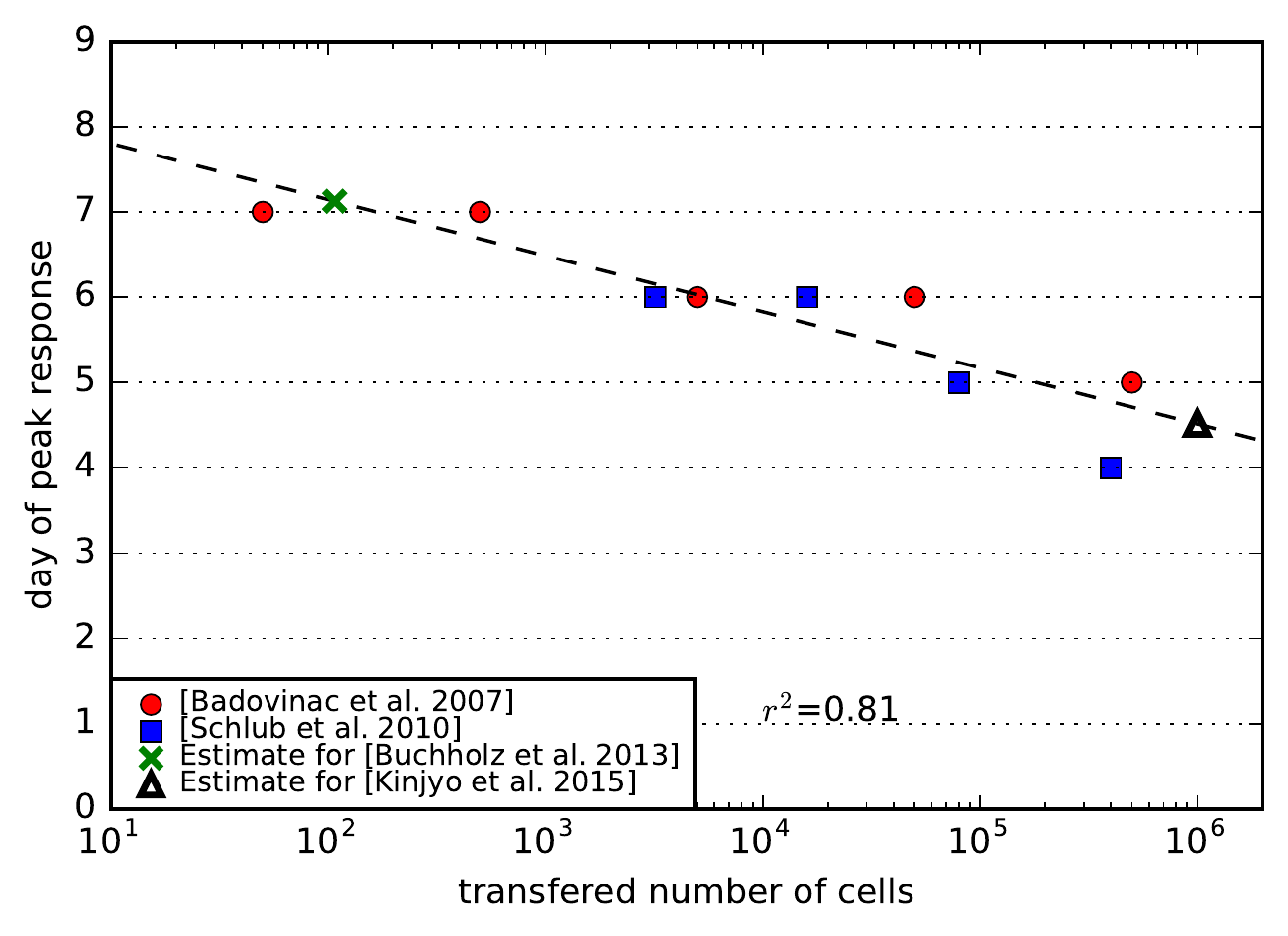}\label{fig:5a}
		\end{subfigure} 
		\begin{subfigure}[t]{0.5\textwidth}
		\caption{.}
		\includegraphics[width=\textwidth]{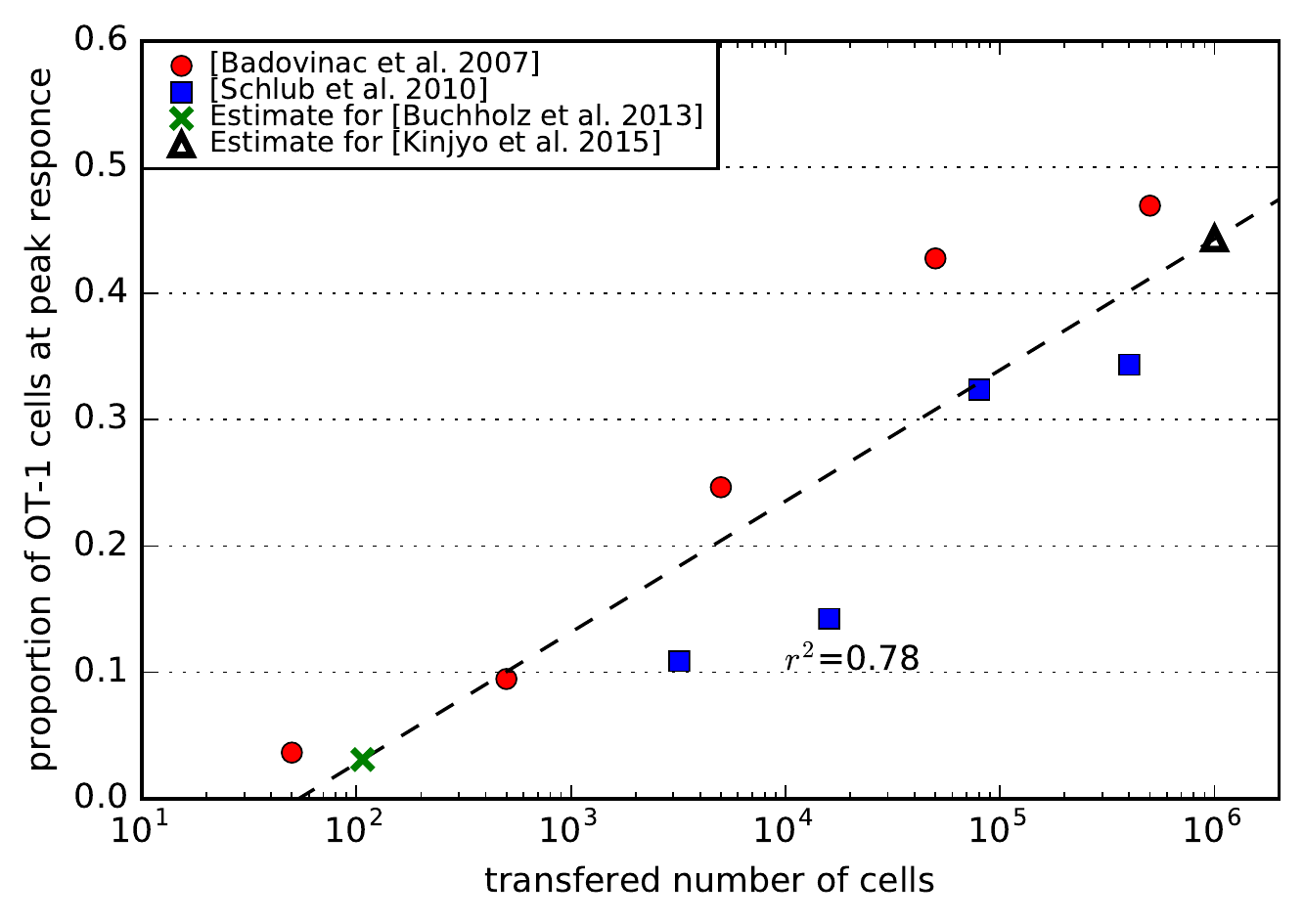}\label{fig:5b}
		\end{subfigure} 
		\begin{subfigure}[t]{0.5\textwidth}
		\centering \caption{.}
		\includegraphics[width=\textwidth]{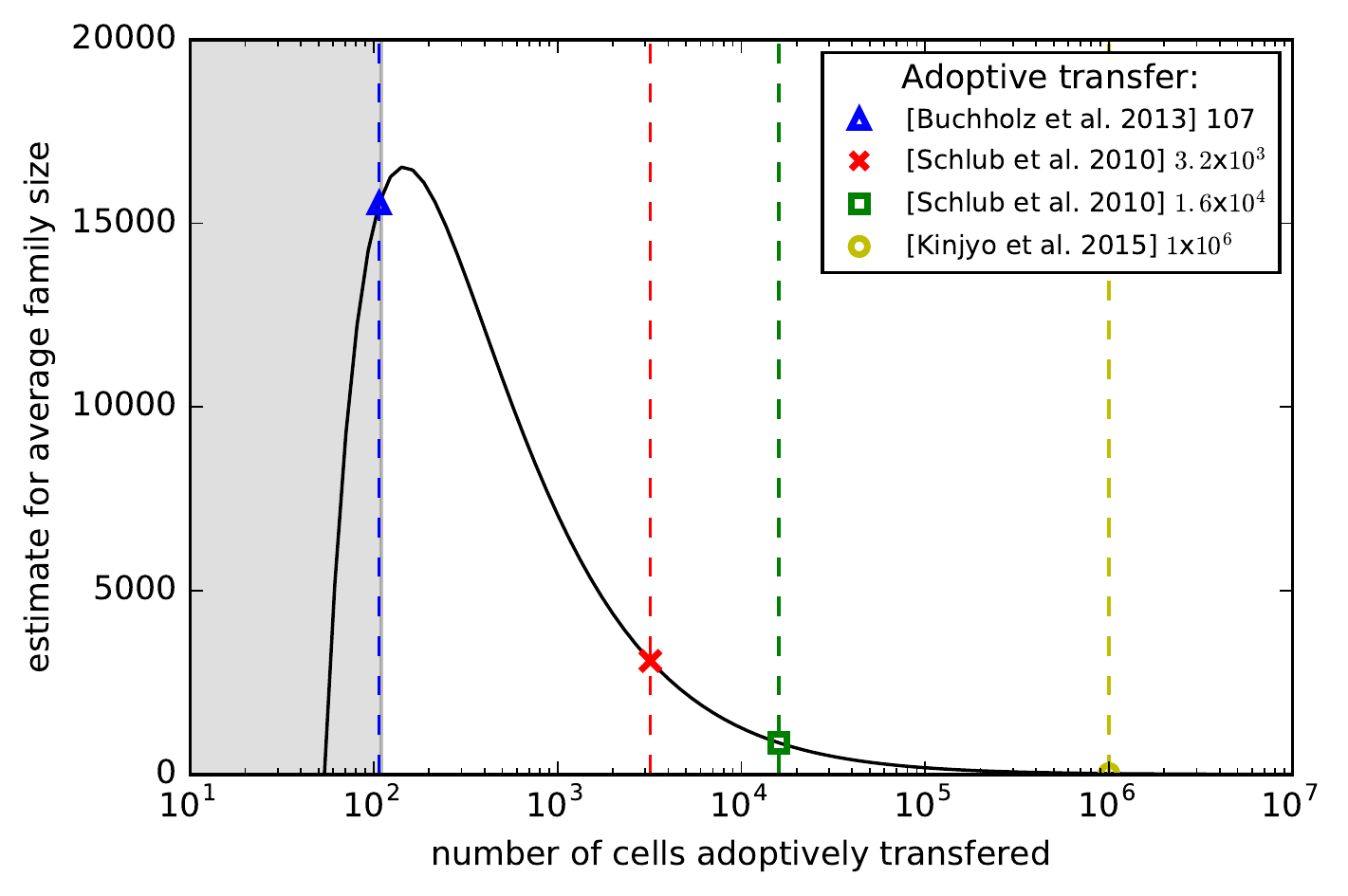}\label{fig:5c}
		\end{subfigure} 
		\caption{\textbf{
		Estimates of the day of peak response and
                average family size as a function of the number of
                adoptively transferred cells, using
		data from \cite{Badovinac2007,SchlubMain}.}
		(A) Log-linear
		regression was used to find an estimate for the day
		of peak response (y co-ordinate) from
		the number of adoptively transferred cells (x
		co-ordinate) reported in \cite{Kinjyo2015} and
		\cite{Buchholz630}. (B) as in A but estimating
		proportion of OT-1 cells of lymphocytes at peak.
		(C) Estimate of average family size at peak for given number
		of adoptively transferred cells. For low numbers of transferred
		cells ($<$100) the estimate is not good.
		} \centering
	\end{figure}

	It is estimated that a mouse has 100--1000 cells capable of
	responding to any specific antigen, meaning large adoptive
	transfers such as $10^6$ used in \cite{Kinjyo2015} dwarf
	the endogenous response \cite{Badovinac2007}. The papers
	\cite{Badovinac2007,SchlubMain} establish that
	increased transfer size leads to an earlier peak response.
	We used log-linear fitting to approximate the relationship
	between the number of transferred cells and the day of
	peak response, Fig. \ref{fig:5a}. For an experiment
	transferring 107 cells, as described in \cite{Buchholz630},
	we estimate the peak to be at day between seven and eight.
	For an experiment transferring $10^6$ cells, as described
	in \cite{Kinjyo2015}, we estimate the peak day to be between
	day four and five. We proceed with the assumption in
	\cite{Buchholz630} that the expansion phase in that system continues 
	until day eight.

		 \label{sec:AvgFamSize} 

	The $\chi^2$ objective function employed by \cite{Buchholz630}
	uses estimates of the sample variance, $\hat{\sigma}_i^2$, for
	each observed statistic. There is not, however, sufficient
	data reported to make the bootstrapping calculation for
	the SEM of \cite{Kinjyo2015} data. As a result we needed to modify the weighting in the 
	objective function so it was not dependent on this missing
	information. This weighting also has to manage the fact
	that an average family size is on a different scale than
	a proportional statistic. Let us suppose we have an observed average
	family size denoted as $\hat{y_1}$ and $k$ statistics observed of
	proportional phenotypes at different times denoted as
	$\hat{y}_2,...\hat{y}_{k+1}$. The objective function we applied was
	a weighted mean squared error defined by
	\begin{align*}
	d_3(\hat{y},\vec{y}(\theta)) =\frac{(\hat{y_1} - y_1(\theta))^2}{\hat{y}_1^2} +\sum_{i=2}^k(\hat{y_i} - y_i(\theta))^2.
	\end{align*}
	That is, we use another weighted mean squared error where we
	only divide one of the sum's elements by the respective
	observed statistic squared, the observed populations at day
	eight post infection. This weighting is required to ensure that the population statistic does not disproportionately influence the objective function, as the population scales exponentially over time while the proportional statistics do not.

	The methodology of fitting models to observed cohort data
	requires an estimate of the average family size at peak
	immune response in order for models to scale correctly,
	however this data is not reported in \cite{Kinjyo2015}. The
	authors of \cite{SchlubMain} provide a methodology for
	estimating the difference in the average number of divisions
	transferred OT-1 CD8+ T cells have made by the time of peak
	response between two experiments. We adopt and extend this
	method to estimate average family size at peak. We fit a
	log linear relationship to estimate the percentage of OT-1
	cell at peak of the immune response for different adoptive
	transfer amounts, Fig. \ref{fig:5b}. We also make the
	assumption that the total number lymphocytes (OT-1 and endogenous) at the peak
	immune response is the same across all experiments, an
	assumption supported by spleen data in \cite{SchlubMain}.
	Spleen counts reported in \cite{Badovinac2007} suggest
	this may not always be the case, particularly for low
	transfer numbers. Finally we assume that the average family
	size reported in \cite{Buchholz630} is representative. Combing
	these assumptions allowed us to derive the average family
	size for a given adoptive transfer of OT-1 cells, Fig.
	\ref{fig:5c}. The log-linear assumption is used for
	simplicity, but predicts negative growth for low numbers
	of OT-1 cells, and so can only be useful above a threshold.

	This scheme indicates that with a transfer of $1\times 10^2$
	cells, as in \cite{Buchholz630}, each clone results in 
	an average of nine more divisions than the clones in a
	transfer of $1\times10^6$ cells, as in \cite{Kinjyo2015}.
	This scales roughly with that estimated in \cite{SchlubMain}
	that a transfer of $3.2\times10^3$ cells results in ~five more
	divisions than a $4\times10^5$ transfer.  We calculate that
	in the experiment reported in \cite{Kinjyo2015} we would
	expect 23 cells per family (four to five average divisions)
	at the peak of around day 4.5. This is reasonable when
	compared to data in \cite{Badovinac2007}, which reported
	that there is only a 13 fold increase in total OT-1 proportion at
	peak for a $10^4$ fold increase in precursor transfer.
	While relative numbers are comparable, large discrepancies
	in reported data for absolute cell counts highlight caveats
	with these estimates. On transferring $10^2$ cells,
	\cite{Buchholz630} reports an average peak family size of
	15k, while \cite{Badovinac2007} gives a 400k estimate for 50 cells
	transferred. Our estimate for the total number of lymphocytes
	in a mouse, calculated from the value reported in \cite{Buchholz630},
	is four fold lower than that \cite{SchlubMain} reported in
	the spleen. Because of the fragility of this calculation,
	we performed a sanity check which established there was
	no significant impact (data not shown).

\begin{figure}[h]
		\begin{subfigure}[t]{0.5\textwidth} \caption{.}
		\includegraphics[width=\textwidth]{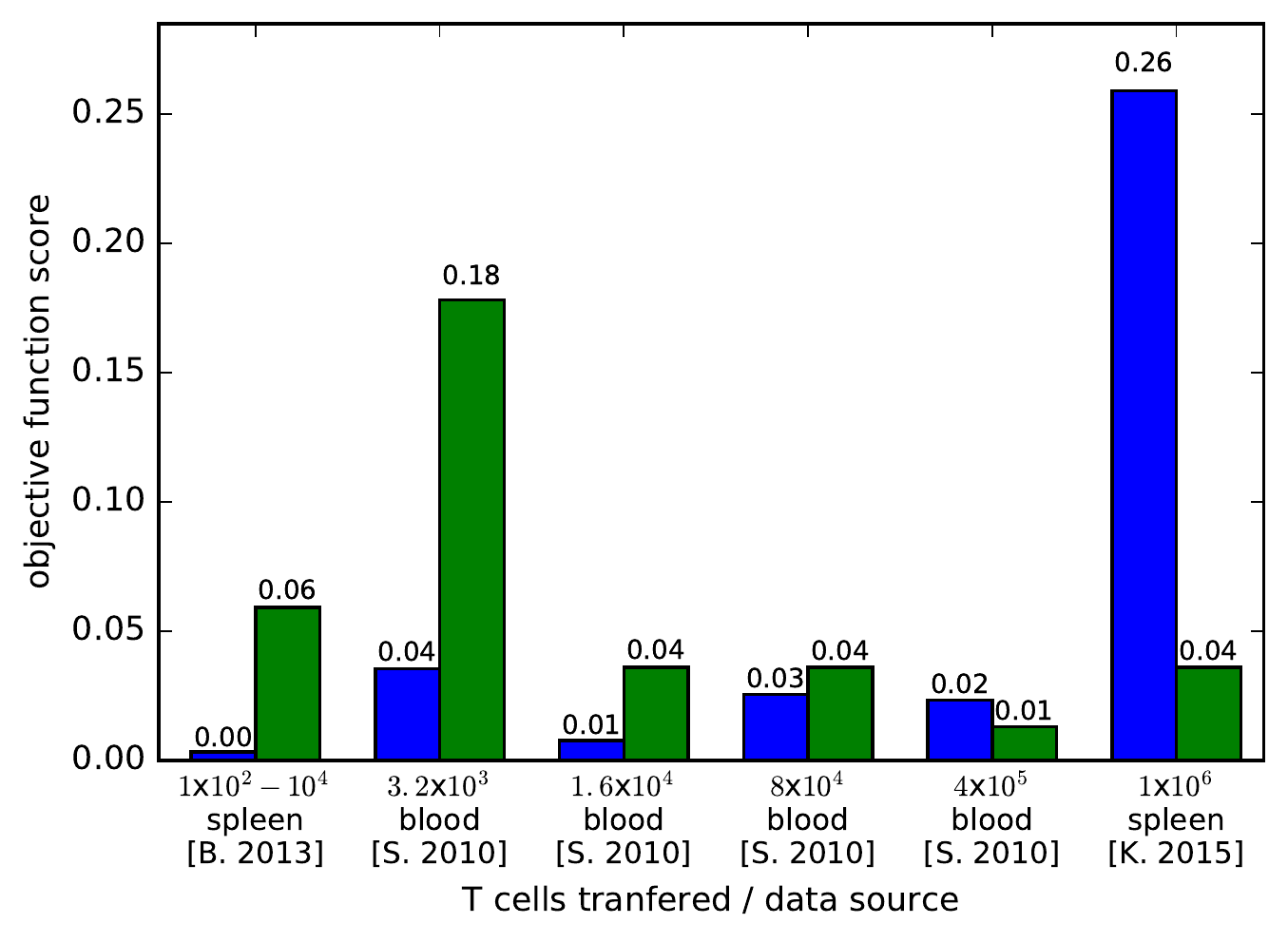}\label{fig:6a}
		\end{subfigure} \begin{subfigure}[t]{0.5\textwidth}
		\caption{.}
		\includegraphics[width=\textwidth]{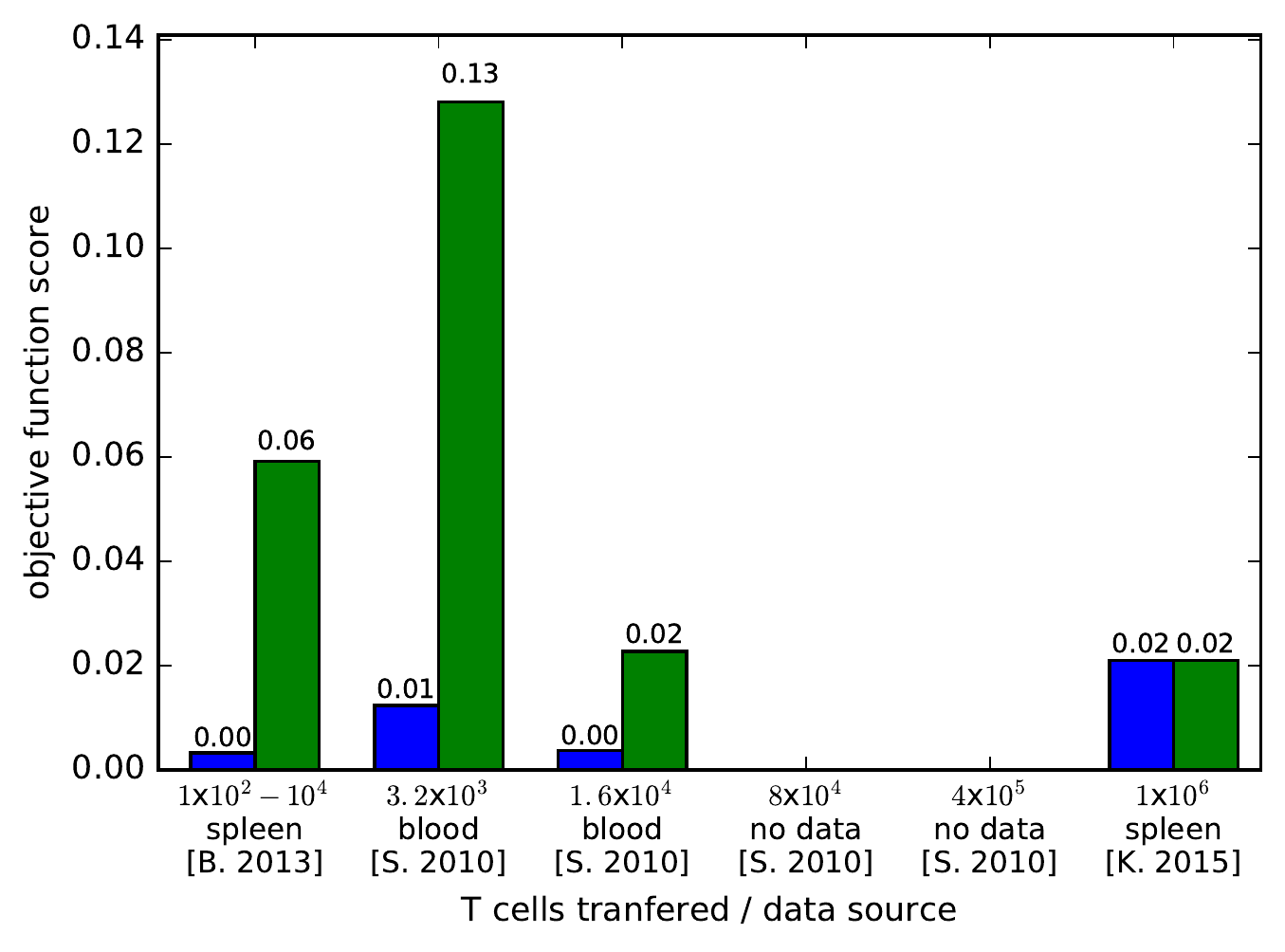}\label{fig:6b}
		\end{subfigure} \begin{subfigure}[t]{0.5\textwidth}
			\caption{.}
			\includegraphics[width=\textwidth]{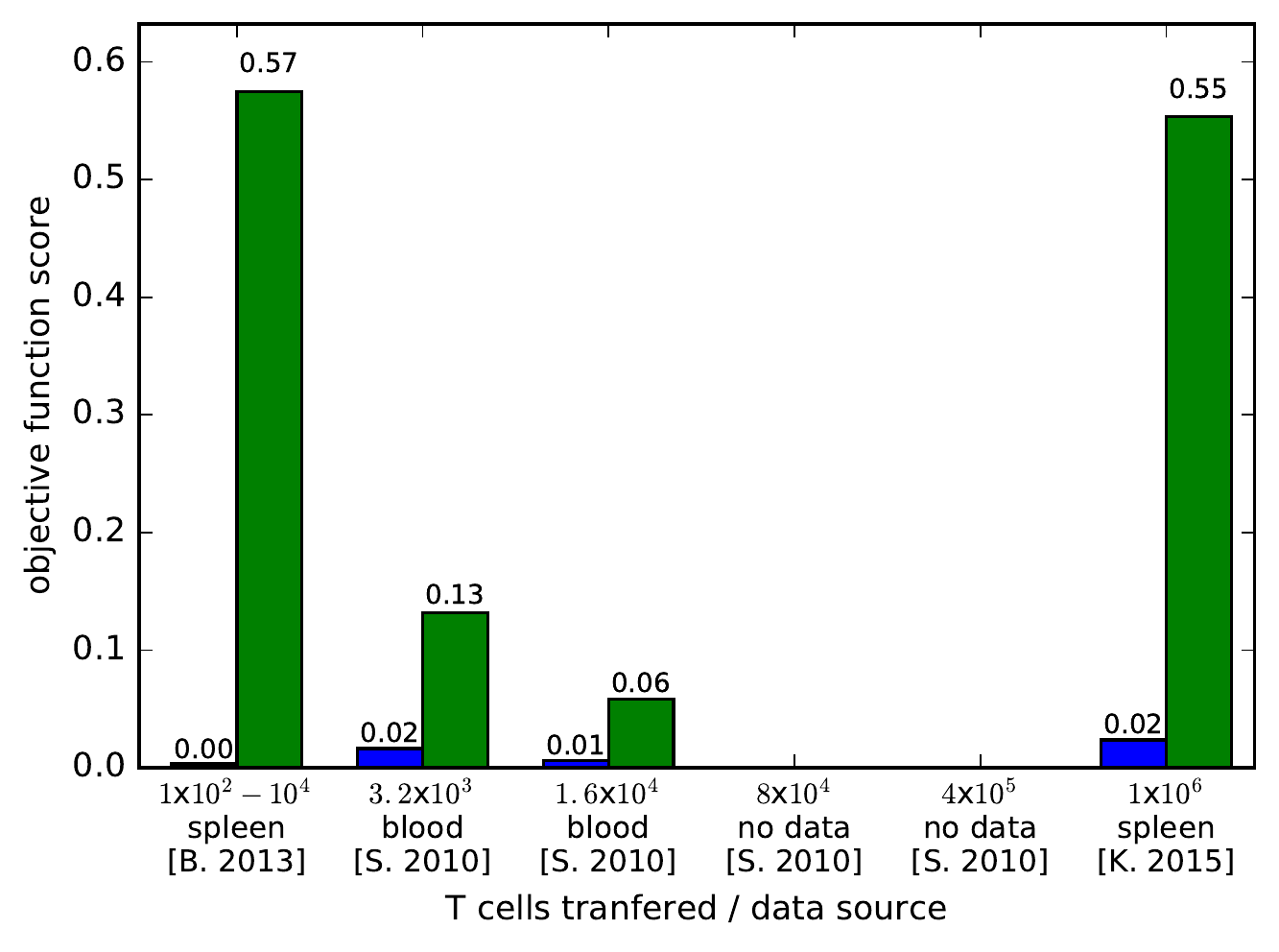}\label{fig:6c}
		\end{subfigure} \begin{subfigure}[t]{0.5\textwidth}
			\caption{.}
			\includegraphics[width=\textwidth]{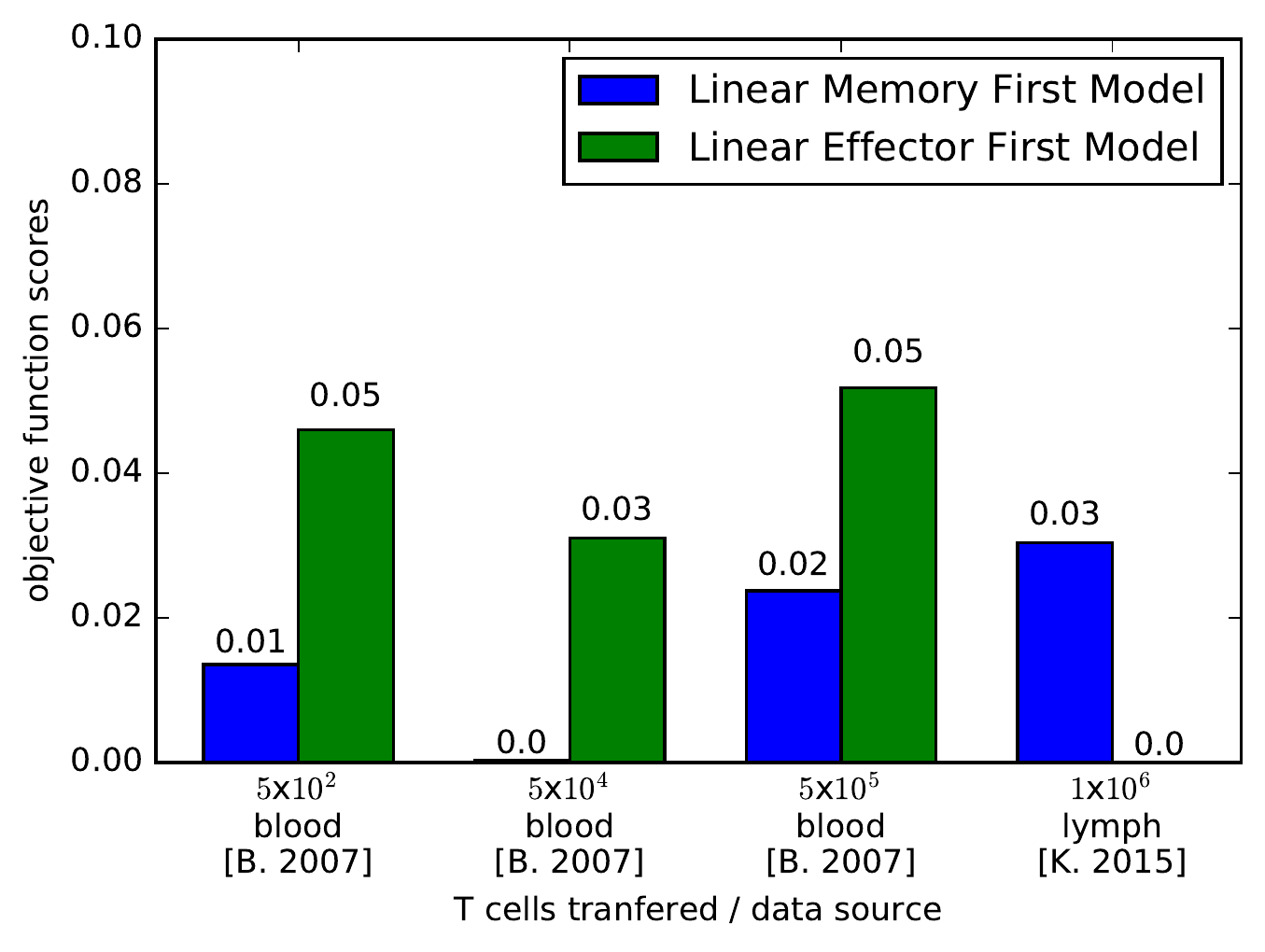}\label{fig:6d}
		\end{subfigure} \caption{ \textbf{The objective
		function value for the two simplified linear models for
		the best parameterizations
		when fitting cohort statistics and average family
		size.}
		Smaller numbers indicate a better fit. Fits only
		done when data sets had four or more data points.
		(A) Fitted to all reported data up to day eight
		post infection. (B) As in A, but curtailing 
		data to the estimated expansion phase. (C) As in B, but with
		the average cell lifetime limited to being below five days. (D)
		as in C, but fit to additional data sources. }
	\end{figure}

	\textbf{Application.} When fitting to all the data in
	\cite{Kinjyo2015} up to day eight post infection, contrary
	to the deduction in \cite{Buchholz630}, the method indicated that
	the Linear Effector First Model provided the better fit, Fig.
	\ref{fig:6a}. The Linear Effector First Model was also the
	best fit for the experiments in \cite{SchlubMain} with high
	adoptive transfer numbers. For experiments with lower numbers
	of adoptively transferred cells, the deduction flipped and
	the Linear Memory first model was the best fit. Thus,
	when fitting without adjusting for the number of cell
	adoptively transferred, we found contradictory results.

	We restricted the data from these papers to the estimated
	expansion phase.  Curtailed data with less than four data
	points was not considered. For all other data sets, the
	Linear Memory First Model provided the best fit, Fig.
	\ref{fig:6b}. On examination, however, all of best fits, 
	had biologically implausible parameterizations, with naive
	cell average lifetimes taking values far over 10 days, with the one exception of the Memory First model fitting to \cite{Buchholz630} data. This is at odds
	with data showing that, even for high adoptive transfers,
	less than 0.1\% of the OT-1 population comprised unrecruited
	cells \cite{SchlubMain}, and data from \cite{Kinjyo2015}
	showing that while no cells had divided on day two post
	infection the majority were preparing to divide. We therefore
	further adapted the method, re-performing the analysis,
	but upper bounding parameters so that average cell lifetime
	is five days or less. When fitting with this adjustment,
	the Linear Memory First Model still provided the best fit
	to all the curtailed data sets, but the result was stronger,
	with the Linear Effector First Model failing to give a good
	explanation for the data in \cite{Buchholz630, Kinjyo2015},
	Fig. \ref{fig:6c}. This result suggested that under the
	assumptions of the model there was no biologically plausible
	good fit for the Linear Effector First Model when fit to
	either data set. We applied the fitting method to blood
	data from \cite{Badovinac2007} which was not included in
	the main analysis, and this also returned a Memory First
	model as the best fit, Fig. \ref{fig:6d}. We also applied
	the method to MLN data reported in \cite{Kinjyo2015}, which,
	in contrast to the other data, returned the Linear Effector
	First Model as the best fit. The MLN data, however, may not
	be appropriate for this analysis, due to the high levels
	of migration.

\begin{figure}[h]
		\begin{subfigure}[t]{\textwidth} \centering
		\caption{.}
		\includegraphics[width=\textwidth]{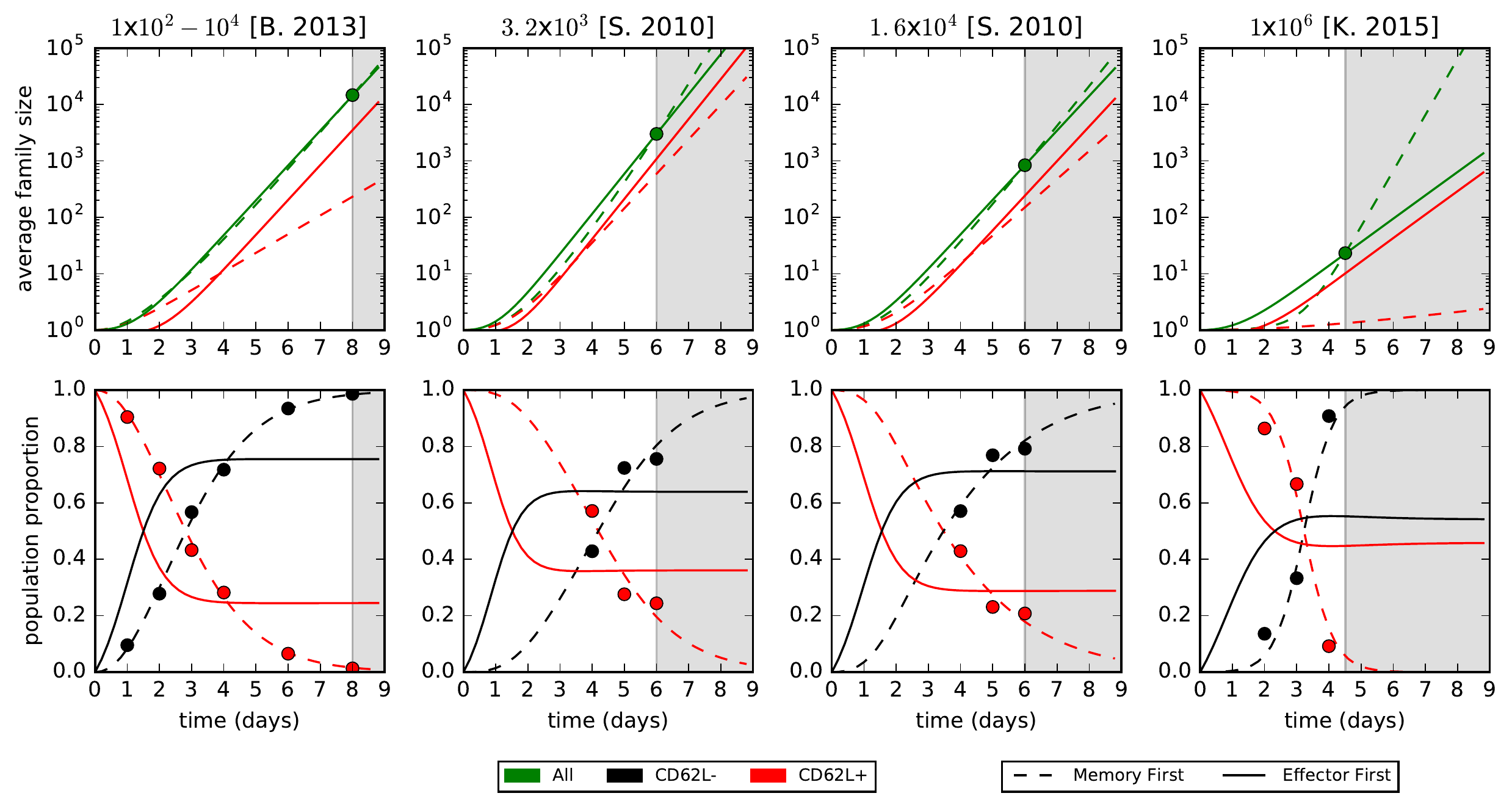}\label{fig:7a}
		\end{subfigure} \begin{subfigure}[t]{0.5\textwidth}
			\caption{.}
			\includegraphics[width=\textwidth]{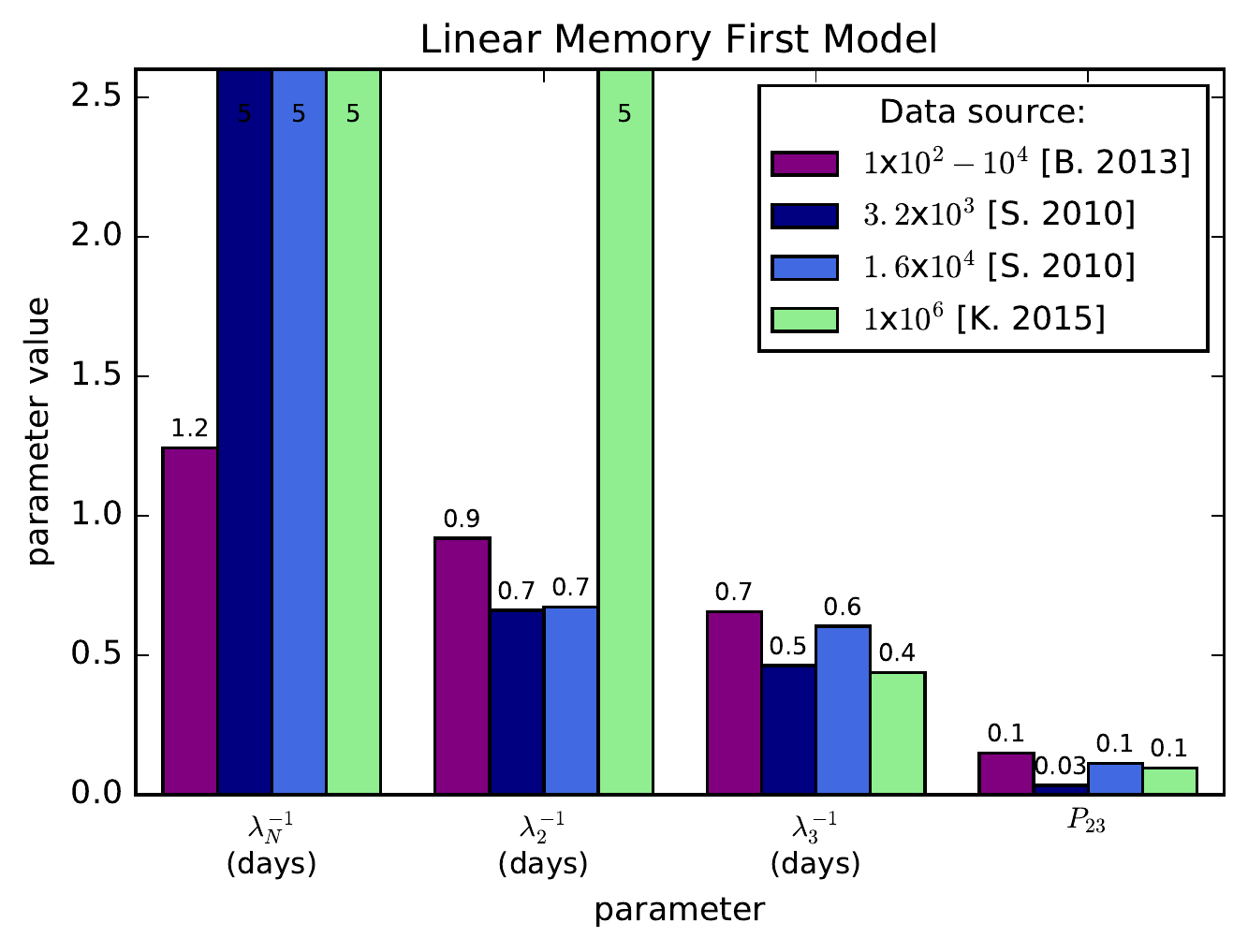}\label{fig:7b}
		\end{subfigure} \begin{subfigure}[t]{0.5\textwidth}
			\caption{.}
			\includegraphics[width=\textwidth]{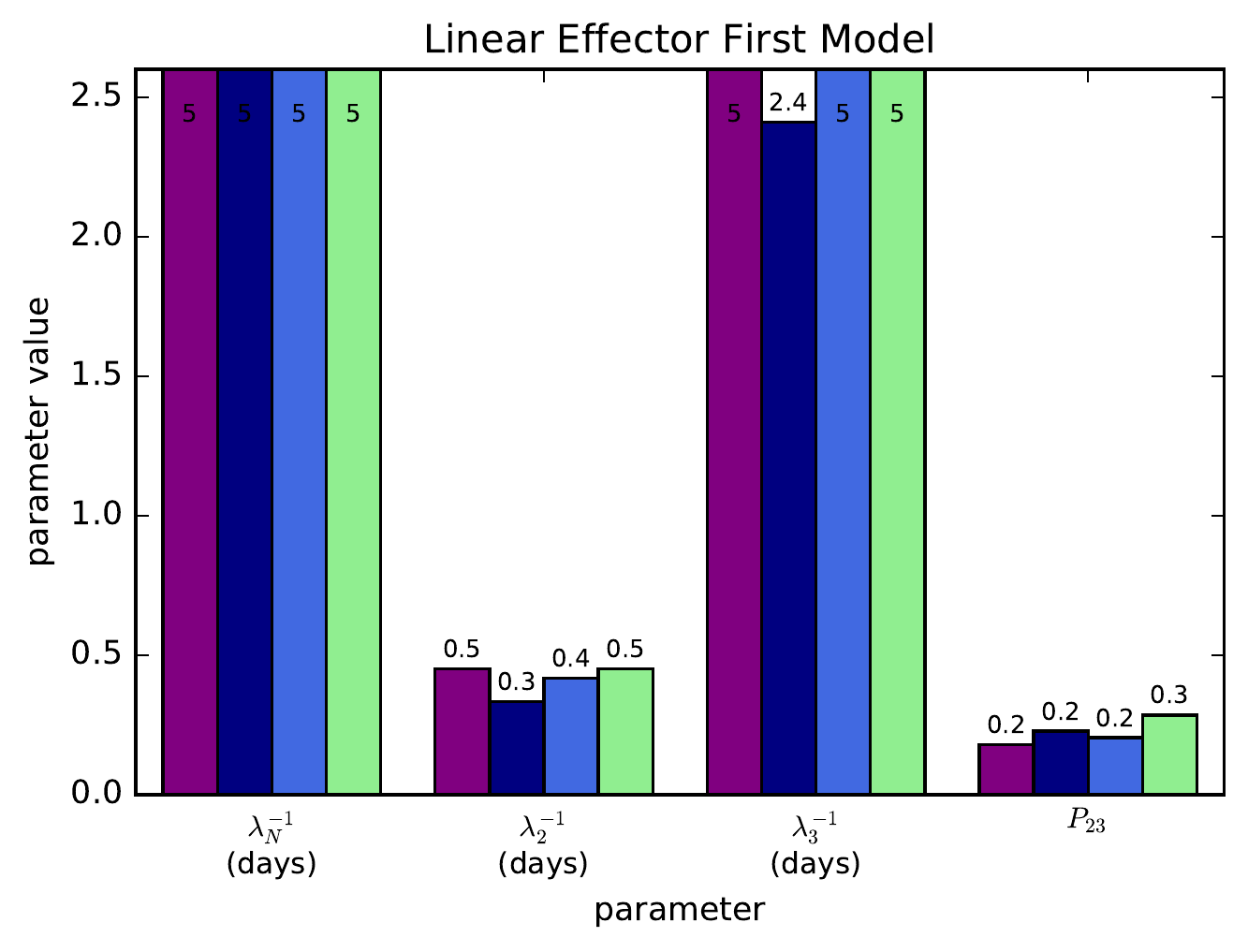}\label{fig:7c}
		\end{subfigure} \caption{ \textbf{Statistics and
		parameterization for the best fitting models for
		cohort data from \cite{Buchholz630, Kinjyo2015,
		SchlubMain} as in Fig. \ref{fig:6c}.} (A)
		Dashed lines show the Linear Memory First Model
		predicted statistics, solid lines show the Linear
		Effector First Model predicted statistics, dots
		show experimental data fitted to and the grey area shows
		the estimated time after the expansion phase has ended. (B)
		Parameters for the best fit Linear Memory First Model.
		(C) Parameters for the best fit Linear Effector First Model.}
		\label{fig:7}
	\end{figure}

	The method chose boundary values for at least one parameter
	in all cases except when fitting the Linear Memory First
	Model to \cite{Buchholz630} data, Fig. \ref{fig:7}. In
	addition, some models predicted little memory, primarily
	fitting the CD62L\textsuperscript{+} portion with naive and
	effector cells, as was the case with the Effector First
	Model fitted to \cite{Kinjyo2015}. This suggested that the
	model may not be sophisticated enough or did not have enough
	data to fit to. An alternative explanation could
	 come from other \cite{Kinjyo2015} results, which showed a
	 homogeneous population effectors on day four post infection
	 make way for two homogeneous groups of effectors and memory
	 cells on day seven. With the insight that day seven is
	 after the expansion phase in the experimental setup described
	 in \cite{Kinjyo2015}, this result suggests the possibility
	 that memory may appear after the strictly exponential
	 expansion phase considered by the model in \cite{Buchholz630}
	 has come to an end.

	  

	\section{Discussion} 
	The evidence presented here suggests that the modeling
	methodology introduced in \cite{Buchholz630} to discriminate
	between differentiation pathways, originally implemented
	to fit to clonal data, is robust in its deductions when
	modified to fit to more readily available time-course
	non-clonal data. Robustness is a key advantage of the
	approach for quickly evaluating differentiation networks.

	We acknowledge that our application of the model to published
	data has several limitations beyond the original: we
	considered a much more limited collection of linear-only
	paths; we had to collapse phenotypic definitions due to not
	having CD27 measurements; we extracted data from papers
	directly from graphs; we employed an adapted method for
	estimating day of peak immune response, and average family
	size.

	Despite all of those considerations, the model provided a
	robust outcome when fit to the non-clonal cohort time-course
	published in \cite{Buchholz630}. Once day of peak response
	was taken into account, it produced consistent deductions,
	that within this modeling framework Memory First provides
	the best fit, across a range of data from other papers
	\cite{Badovinac2007,SchlubMain,Kinjyo2015}. When inappropriately
	fit to time-courses that extend beyond the expansion phase,
	interestingly it supported the Effector First Model in some
	cases, which is more consistent with other data in
	\cite{Kinjyo2015}.

	As with any modeling framework, there are, of course, 
	caveats in the strength of the conclusions drawn from it.
	Presumably due to the inherently limited nature of the
	\emph{in vivo} data being fit, several modeling assumptions
	were made in \cite{Buchholz630} that keep the
	computational model tractable and parameters identifiable,
	but some of which are inconsistent with published experimental
	data. As more sophisticated tracking of cell familial fates
	and information on physiological constraints around lineage
	transitions develops, we anticipate that ultimately other
	mathematical models will be employed that are built on that
	knowledge. For illustration, we attempt to highlight where
	those assumptions may influence the deductions of the present
	study.

	The model's remit is exclusively an expansion phase with
	no death and strictly exponentially growing populations.
	As a result, the method cannot be reasonably expected to
	draw inferences on differentiation if it occurs as motivation
	to divide is petering out.

	The fundamental unit of autonomous randomness in the model
	is the cell: each cell is born afresh, independently selecting
	its lifetime and fate in a cell-type dependent fashion. The
	data in both \cite{Buchholz630,Gerlach635} illustrate a
	strong familial influence, with some clones expanding
	dramatically more than others. Controlled \emph{in vitro}
	experiments have established significant elements of clonal
	dependency \cite{lemaitre2013phenotypic}, notably in burst
	size \cite{Mar2016,PhilH1}, while \emph{in vivo}
	data using the cell cycle reporter FUCCI
	\cite{sakaue2008visualizing} has established that during
	an adaptive immune response cells drop out of cycle 
	as early as day four and the number of non-dividing cells
	progressively increases with time \cite{Marchingo1123}.
	Thus, to retain simplicity and identifiability, an alternative
	view, taking to account these experimental features would
	posit that the ultimate clone burst size is randomized
	between families, and that the slowing in proliferation
	rate noted by the memory first model, Fig. \ref{fig:4b}
	and Fig. \ref{fig:7b}, is created by non-dividing memory
	cells diluting the average division time of the OT-1
	populations being measured.

	The use of an exponential distribution for the lifetime
	distribution ensures that the multi-type Bellman-Harris
	process is Markovian, and facilitates the use of non-asymptotic
	values in model fitting. Due to the minimum time needed
	to synthesize DNA, there is always a lower-bound on division
	times and it has been long since known that for lymphocytes,
	at least \emph{in vitro}, the exponential does not offer a
	good description of the distribution of division times and
	a right skewed distribution is more appropriate
	\cite{smith1973cells, Dowling29042014, Hawkins20032007,
	Hawkins11082009, Duffy338}. Thus, while cohort and clonal data yielded the same deductions in our analysis, only experiments looking deeper than the population level can shed light on the validity of the modeling assumptions used for inference.

	Whether a model built taking these assumptions into account
	would lead to distinct deductions is unclear. What is clear
	is that, within the confines of the assumptions underlying
	it, the model introduced in \cite{Buchholz630} provides a
	method for quickly evaluating differentiation networks,
	even for cohort level data.

	{\bf Acknowledgement.}
	The work of A.M. and K.R.D. was supported by Science
	Foundation Ireland Grant 12IP1263. This work was supported
	by the National Health and Medical Research Council of
	Australia via Project Grant 1010654, Program Grant 1054925
	and a fellowship to P.D.H. as well as an Australian Government
	National Health and Medical Research Council Independent
	Research Institutes Infrastructure Support Scheme Grant
	361646.

	\bibliography{Report_bibliography} 
	\bibliographystyle{plain}
\end{document}